\documentstyle[12pt]{article}
\newcommand{\bd}{\begin{displaymath}}
\newcommand{\ed}{\end{displaymath}}
\newcommand{\be}{\begin{equation}}
\newcommand{\ee}{\end{equation}}
\newcommand{\barr}{\begin{eqnarray}}
\newcommand{\earr}{\end{eqnarray}}
\newcommand{\barrr}{\begin{eqnarray*}}
\newcommand{\earrr}{\end{eqnarray*}}
\newcommand{\da}{\delta a}
\newcommand{\db}{\delta b}
\newcommand{\dh}{\delta h}
\newcommand{\dv}{\delta v}

\title{Direction dependent free energy singularity\\
of the asymmetric six-vertex model}
\author{\em Giuseppe Albertini\thanks{E-mail address: 
albert@osfmi.mi.infn.it}\\
\em Physics Dept. of Milan University\\
\em via Celoria 16, 20133 Milano (Italy)} 
\date{}

\begin{document}

\maketitle

\begin{abstract}
The transition from the ordered commensurate phase to the
incommensurate gaussian phase of the antiferroelectric asymmetric
six-vertex model is investigated by keeping the temperature constant
below the roughening point and varying the external fields
$(h,v)$. In the $(h,v)$ plane, the phase boundary is approached
along straight lines $\delta v=k \delta h$, where $(\delta h,\delta v)$
measures the displacement from the phase boundary. It is found that
the free energy singularity displays the exponent $3/2$ typical of
the Pokrovski-Talapov transition $\delta f \sim \mbox{const} 
(\delta h)^{3/2}$ for any direction other than the tangential one.
In the latter case $\delta f$ shows a discontinuity in the third
derivative.
\end{abstract}

\newpage

\section{Introduction}
The asymmetric six-vertex model is the extension of
the well-known symmetric six-vertex model to the case
where external fields interact with the local fluctuating
variables (``dipoles'' or ``arrows'') \cite{SYY,No,Ga}. On a two-dimensional
square lattice, two-valued variables live on links and interact at
vertices, and, throughout this paper, we choose interactions
that favor antiferroelectric order. An external homogeneous
two-component field $(h,v)$, favoring ferroelectric ordering,
competes with the arrow-arrow couplings.

If one treats the problem with the transfer matrix method, an
exact solution is provided by the Bethe-ansatz \cite{SYY,No,Ga}. The phase
diagram was outlined in the original paper \cite{SYY}, while further
developments came more recently \cite{No,NK,K,ADW1,ADW2} 
(the list is not exhaustive.
Papers dealing with the ferroelectric regime are not included).
Since the early works appeared, it has been known that, in the $(h,v)$ 
plane, there is a closed
curve $\Gamma$ separating an antiferroelectrically ordered
(commensurate) massive region, inside $\Gamma$, from an
incommensurate, disordered, massless region, outside $\Gamma$.
The ordered region includes the symmetric model point $(0,0)$
and the free energy in it is field independent
\bd
f(\gamma,u,h,v)=f_{0}(\gamma,u)
\ed
The parameters $\gamma, u$ contain the temperature dependence
and will be defined in the next section. As $\Gamma$ is 
approached from outside (i.e. from the massless phase), keeping
$\gamma$ and $u$ fixed, the free energy is expected to display 
a singularity with a characteristic exponent $3/2$. More precisely,
the following result has been proven in \cite{LW}.

Let $(\overline{h}(b),\overline{v}(b))$ be a parametric
equation of $\Gamma$, to be given later, with $b$ some real
parameter running in a bounded interval. Lieb and Wu found
that, if $h=0$, which implies $b=0$, the singular part of the
free energy is
\be\label{p1}
f(\gamma,u,0,\overline{v}(0)+\delta v)=f_{0}(\gamma,u)+
\delta f(\gamma,u,\delta v)=f_{0}(\gamma,u)+c(\gamma,u)
(\delta v)^{3/2}
\ee
yielding a divergence $\sim \delta v^{-1/2}$ for the susceptibility.
(A divergence with exponent $1/2$ was also found for the specific
heat at fixed field). Borrowing a terminology introduced in later
years, the transition belongs to the Pokrovski-Talapov universality
class \cite{PT,dN}.

Eq.(\ref{p1}) does not clarify how the singularity depends on both field
components. A further step in this direction was recently achieved
by Noh and Kim \cite{NK} who extended the method of \cite{BIK}
to relate macroscopic
quantities such as susceptivities to finite size corrections. They
showed that, in the critical phase
\be\label{p2}
\left|
\begin{array}{ll}
\frac{\partial^2 f}{\partial h^2} & \frac{\partial^2 f}
{\partial h \partial v} \\
\frac{\partial^2 f}{\partial v \partial h} & 
\frac{\partial ^2 f}{\partial v^2}
\end{array}
\right|= (\frac{2}{\pi g})^2
\ee
where $g$ is the coupling constant of the gaussian model on which the 
critical incommensurate phase renormalizes. As $\Gamma$ is approached 
from this phase, $g \rightarrow 2$.

In this paper, the question is settled of how, at fixed $\gamma, u$ the 
singularity of $f(h,v)$ depends on both $(h,v)$. While finding the 
leading singularity as a function of $(h,v)$ simultaneously has proven 
to be elusive, it will be shown that the singularity does depend on the 
direction in the $(h,v)$ plane.

The results are best summarized if one introduces the following notation. Set
\barrr
v_{t}(b) &=& \frac{d}{db} \overline{v}(b) \ \ \ \ h_{t}(b)=\frac{d}{db}
\overline{h}(b) \\
v_{1}(b) &=& \frac{d}{db}v_{t}(b) \ \ \ \ h_{1}(b)=\frac{d}{db}v_{t}(b) \\
\Delta  &=& v_{t}h_{1} - h_{t}v_{1}
\earrr
Next, consider a variation into the incommensurate phase $h=\overline{h}(b)+
\delta h$, $v=\overline{v}(b) + \delta v$ where $\delta v = k \delta h$ and
$k$ fixes a slope not tangential to $\Gamma$. Then
\be\label{p3}
\delta f (\delta h) = (\frac{h_{t}}{\Delta})^{1/2} \frac{h_{t}}{3\pi}
[2(k-\frac{v_{t}}{h_{t}})\delta h]^{3/2}
\ee
so the singularity is governed by an exponent $3/2$ for all these directions.
Yet, if $\Gamma$ is approached tangentially, which amounts to take
$\delta v = \frac{v_{t}}{h_{t}} \delta h $
\be\label{p4}
\delta f(\delta h) = \left\{ \begin{array}{l} 
c_{+}\delta h^3\ \ \ \mbox{if}\ \ \  \delta h > 0 \\
c_{-}\delta h^3\ \ \ \mbox{if}\ \ \  \delta h < 0 
\end{array} \right.
\ee
where $c_{+} \neq c_{-}$ are $b$-dependent and are given in Eq. (\ref{p38}).
In other
words there is a jump in the third derivative.
The calculation breaks down an the two points on $\Gamma$ where $h_{t}=0$, 
i.e. where the tangent to $\Gamma$ is parallel to the $v$-axis. Those cases 
were examined in \cite{ADW2} and an analogous conclusion was reached.

The paper is divided as follows. Section 2 presents a summary of the
Bethe-ansatz and already known results. Section 3 deals with a
perturbative expansion of the integral equations typical of the Bethe-ansatz
and section 4 exploits that expansion to determine the singularities of
the free energy.

\section{Basic definitions and summary of Bethe-ansatz}
The Boltzmann weights of the six allowed configurations are grouped
into $R_{\alpha \alpha'}^{\beta \beta'}(u)$ as shown in Fig.1. In the
framework of the Bethe-ansatz, the
spectral parameter notation, chosen here, seems more natural than the
traditional one. Row-to-row transfer matrices
\bd
T(u)_{\underline{\alpha}, \underline{\alpha'}}=
\sum_{\underline{\beta}} \prod_{k=1}^{N} R_{\alpha_{k} \alpha_{k}'}
^{\beta_{k} \beta_{k+1}}(u)
\ed
with periodic boundary conditions $(\beta_{N+1}=\beta_{1})$ commute
for different values of the spectral parameter
\bd
[T(u),T(u')]=0
\ed
Arrow conservation at each vertex, and periodic boundary conditions,
imply that $T(u)$ breaks into blocks between states with the same number of
up (and down) arrows. Let $n$ be the number of arrows reversed with
respect to the reference state $|\uparrow \uparrow, \ldots , 
\uparrow>$. The Bethe-ansatz provides the following solution to the
eigenvalue problem for $T(u)$ \cite{SYY,No,Ga,ADW2}:
\barr\label{p5}
\Lambda(u) &=& e^{v(N-2n)+hN}[\frac{\sinh(\gamma -u)}{\sinh \gamma}]^N
\prod_{j=1}^{n}\frac{\sinh (\frac{\gamma}{2}+u-\frac{i\alpha_{j}}{2})}
{\sinh(\frac{\gamma}{2}-u+\frac{i\alpha_{j}}{2})} \nonumber \\
&+& e^{v(N-2n)-hN}[\frac{\sinh u}{\sinh \gamma}]^N \prod_{j=1}^{n}
\frac{\sinh(-\frac{3\gamma}{2}+u-\frac{i\alpha_{j}}{2})}{\sinh(\frac{\gamma}
{2}-u+\frac{i\alpha_{j}}{2})}
\earr
is an eigenvalue if the ``rapidities'' $\{\alpha_{j}\}$, $j=1,2, \ldots, n$
satisfy the set of equations
\be\label{p6}
[\frac{\sinh(\frac{\gamma}{2}+\frac{i\alpha_{j}}{2})}{\sinh(\frac{\gamma}{2}-
\frac{i\alpha_{j}}{2})}]^N=(-)^{n+1}e^{2hN}\prod_{k=1}^n \frac{\sinh(\gamma
+\frac{i}{2}(\alpha_{j}-\alpha_{k}))}{\sinh(\gamma-\frac{i}{2}(\alpha_{j}-
\alpha_{k}))} \ \ \ \ j=1,2,\ldots ,n
\ee
In the limit $N \rightarrow \infty$, the rapidities $\{\alpha_{j}\}$ condense
into curves in the complex plane, and are conveniently described by a
density function $R(\alpha)$
\bd
R(\alpha_{j})=\lim_{N\rightarrow \infty} \frac{2\pi}{N(\alpha_{j+1}-
\alpha_{j})}
\ed
Eqs. (\ref{p6}) are replaced by a single linear integral equation that
governs the thermodynamics of the model. Introduce the functions
(this notation is somewhat redundant, but it has been adopted in
many previous papers and it will be kept here)
\barrr
p^0(\alpha)&=&-i \log [ \frac{\sinh(\frac{\gamma}{2}+\frac{i\alpha}{2})}
{\sinh(\frac{\gamma}{2}-\frac{i\alpha}{2})} \ \ \
\Theta(\alpha)=-i\log [\frac{\sinh(\gamma + \frac{i\alpha}{2})}
{\sinh(\gamma - \frac{i\alpha}{2})}] \\
\xi(\alpha)&=&\frac{dp^0(\alpha)}{d\alpha} \ \ \ K(\alpha)=\frac{d\Theta
(\alpha)}{d\alpha}
\earrr
and the vertical polarization
\be\label{p7}
y=\lim_{N\rightarrow \infty} (1-\frac{2n}{N})
\ee
Then, for a state described by a rapidity curve $C$, the density
$R(\alpha)$ solves \cite{SYY,No,ADW2}
\be\label{p8}
\xi(\alpha) - \frac{1}{2\pi}\int_{C} d\beta K(\alpha-\beta)R(\beta)=
R(\alpha)
\ee
and
\be\label{p9}
p^0(\alpha)-\frac{1}{2\pi}\int_{C}d\beta \Theta(\alpha-\beta)
R(\beta)+2ih=2\pi x \ \ \ \ \ -\frac{1-y}{4}\leq x \leq \frac{1-y}{4}
\ee
where $x$ is the real parameter of the curve. Let $A=-a+ib$, $B=a+ib$
be the two endpoints of the curve. (We take for granted that
$B=-A^*$ because we wish to consider, in each sector of
fixed $n$, the largest transfer matrix eigenvalue, which is
real and unique by Perron-Froboenius theorem \cite{Gt}. Since $\{\alpha_{j}\}
\rightarrow \{-\alpha_{j}^{*}\}$ is a symmetry of (\ref{p6}) we expect 
it to hold 
for the solution corresponding to the unique largest eigenvalue in each
sector). Solution of (\ref{p8}) implicitly depends on the endpoints
$A,B$ and contributes to make $y,h$ dependent on $a,b$ through \cite{SYY,
No,ADW2}
\be\label{p10}
\frac{1-y}{2}=\frac{1}{2\pi}\int_{A}^{B}d\alpha R(\alpha;A,B)
\ee
\be\label{p11}
p^0(A)+p^0(B)-\frac{1}{2\pi}\int_{A}^{B}d\beta R(\beta;A,B)
[\Theta(A-\beta)+\Theta(B-\beta)]+4ih=0
\ee
In the transfer matrix formalism, the free energy is determined
by the largest eigenvalue $\Lambda_{0}$,
so, neglecting a factor $\frac{1}{k_{B}T}$
\be\label{p12}
f(u,\gamma,h,v)=-\lim_{N\rightarrow \infty} \frac{\ln \Lambda
_{N,0}(u,\gamma,h,v)}{N}
\ee
With an abuse of language, the relevant eigenstate will be called
``ground state''.
From (\ref{p5}), all eigenvalues are such that $\Lambda(u)=\Lambda_{R}(u)+
\Lambda_{L}(u)$. One of the two addends dominates over the others 
for specific values of the parameters. We set
\barr
\lim_{N\rightarrow \infty} \frac{1}{N}\ln \Lambda_{R}(u) &=&
F_{R}(u,\gamma,h,y)+vy= \nonumber \\
&=& h+\ln \frac{\sinh(\gamma-u)}{\sinh \gamma} +
\frac{1}{2\pi}\int_{A}^{B}d\alpha R(\alpha;A,B)f_{R}(\alpha,u) \label{p13} \\
\lim_{N\rightarrow \infty} \frac{1}{N}\ln \Lambda_{L}(u) &=&
F_{L}(u,\gamma,h,y)+vy= \nonumber \\
&=& -h+\ln \frac{\sinh u}{\sinh \gamma}+
\frac{1}{2\pi}\int_{A}^{B}d\alpha R(\alpha;A,B)f_{L}(\alpha;u) \label{p14}
\earr
with
\bd
f_{R}(\alpha;u)=\ln \frac{\sinh(\frac{\gamma}{2}+u-\frac{i\alpha}{2})}
{\sinh(\frac{\gamma}{2}-u+\frac{i\alpha}{2})} \ \ \
f_{L}(\alpha;u)=\ln \frac{\sinh(-\frac{3\gamma}{2}+u-\frac{i\alpha}{2})}
{\sinh(\frac{\gamma}{2}-u+\frac{i\alpha}{2})}
\ed
and call $F=\mbox{max}\{F_{R},F_{L}\}$. The equilibrium polarization and
the free energy are then determined by
\be\label{p15}
f(u,\gamma,h,v)=\min_{-1\leq y \leq 1}(-F(u,\gamma,h,y)-vy)=
\min_{-1 \leq y \leq 1} {\cal F} (u,\gamma,h,v,y)
\ee
The ground state solution of (\ref{p8}) is explicitly computed when
$h$ and $v$ are sufficiently small \cite{SYY,No,Ga,ADW2}. 
It corresponds to $n=N/2$ and,
in the $N \rightarrow \infty$ limit, to $a=\pi$, $-\gamma \leq b \leq \gamma$.
Then (\ref{p8}) can be solved by Fourier transform. Even though the 
solution has
appeared many times in the literature, it is worthwhile to recall it
here to introduce the elliptic function notation
\be\label{p16}
R(\alpha;-\pi+ib,\pi+ib)=\sum_{n=-\infty}^{+\infty}\frac{e^{-in\alpha}}
{2\cosh(n\gamma)}=\frac{I(k)}{\pi}\mbox{dn}(\frac{I(k)\alpha}{\pi};k)
\ee
where $\frac{I'(k)}{I(k)}=\frac{\gamma}{\pi}$ and $I(k) (I'(k))$ is the
complete elliptic integral of the first kind with modulus $k (k'=
\sqrt{1-k^2})$. Replacing (\ref{p16}) into (\ref{p10}) and (\ref{p11}) yields
\cite{SYY,No,Ga,ADW2}
\be\label{p17}
y=0 \ \ \ \ \overline{h}(b) = Z(-b)
\ee
where
\be\label{p18}
Z(x) \stackrel{def}{=} \frac{x}{2} + \sum_{n=1}^{\infty} 
\frac{(-)^n}{n}\frac{\sinh(nx)}{\cosh(n\gamma)}
\ee
Within the range $-\gamma \leq b \leq \gamma$ there is a crossing between
$\Lambda_{R}$ and $\Lambda_{L}$. Let's call $id$ the point where the ground
state curve meets the imaginary axis in the $\alpha$-plane. It turns out
that $\Lambda_{R}$ dominates when $d < \gamma -2u$ while $\Lambda_{L}$
does when $d > \gamma -2u$. One can check it by replacing (\ref{p16}) into
(\ref{p13}),(\ref{p14}), or, more quickly, by the following argument. 
Select chains such that $n=\frac{N}{2}$ is an odd number.
By virtue of the symmetry
$\{\alpha_{j}\} \rightarrow \{-\alpha_{j}^{*}\}$ one $\alpha$, say
$\overline{\alpha}$, has to be pure imaginary. The other $n-1$
$\{\alpha_{j}\}$ can be paired to give a positive contribution to
$\Lambda_{R}$ and $\Lambda_{L}$ so the contribution of $\overline{\alpha}$
determines the sign of $\Lambda_{R}$ and $\Lambda_{L}$. It is easily seen
that
\barrr
&& \Lambda_{R}(\overline{\alpha}) >(<)0 \ \ \mbox{if} \ \
\mbox{Im}\overline{\alpha}<(>) \gamma -2u \\
&& \Lambda_{L}(\overline{\alpha}) >(<)0 \ \ \mbox{if} \ \
\mbox{Im}\overline{\alpha}>(<) \gamma-2u
\earrr
In the thermodynamic limit $\mbox{Im}\alpha \rightarrow d$, and since by
Perron-Froboenius theorem $\Lambda_{N,\mbox{\scriptsize max}}$ must be 
positive, we 
get the desired result. When crossing, though, $\Lambda_{R}$ and
$\Lambda_{L}$ connect smoothly, so no singularity of $f$ appears
and one finds from (\ref{p13}),(\ref{p14}) the field independent value of the
symmetric six-vertex model \cite{Ba}
\be\label{p19}
f(u,\gamma,h,v)=-2 \sum_{n=1}^{+\infty} \frac{e^{-2\gamma n}}
{n\cosh(\gamma n)}\sinh(nu)\sinh n(\gamma -u) \ \ \ 0 \leq u \leq \gamma
\ee
The region in the $(h,v)$ plane where (\ref{p19}) is valid is bounded
by (\ref{p17}) and the value of the $v$-field at which the ground state
moves away from the $y=0$ sector. This is fixed by the equation
\be\label{p20}
\overline{v}(b)= - \frac{\partial F}{\partial y}|_{\mbox
{\scriptsize h fixed, y=0}}
\ee
that yields \cite{SYY,No,Ga,ADW2}
\be\label{p21}
\overline{v}(b)= Z(\gamma - |\gamma -2u-b|) \ \ \ -\gamma \leq b \leq \gamma
\ee
More precisely, $\overline{v}(b)$ in (\ref{p21}) runs over half of the curve,
the other half is obtained from the symmetry $f(h,v)=f(-h,-v)$,
which follows trivially from reversal of all arrows in the
statistical sum. Eqs. (\ref{p17}) and (\ref{p21}) provide the parametric
equation $(\overline{h}(b),\overline{v}(b))$ of $\Gamma$.

\section{The expansion in $\delta a$, $\delta b$}
Eqs. (\ref{p8}), (\ref{p10}), (\ref{p11}), (\ref{p13}), (\ref{p14})
determine the thermodynamics of the system. In
principle, one should solve (\ref{p8}) and plug $R(\alpha;A,B)$
into the others. This would give
\bd
y=y(a,b) \ \ \ \ h=h(a,b) \ \ \ \ F=F(a,b)
\ed
Once the first two are inverted and plugged into the third,
one can minimize $-F(\gamma,u,h,y) - vy$ w.r. to $y$
keeping $h$, $v$ fixed. Of course, (\ref{p8}) is not exactly
solvable for $a \neq \pi$. Therefore we attempt an
expansion that generalizes the method that Lieb and Wu
\cite{LW} applied to the $h=0$ case, where the rapidities are
real $(b=0)$ and everything is a function of $a$ only.
Namely, it will be assumed that a unique solution exists
for (\ref{p8}) at least in a narrow neighborhood of the segment
$a=\pi$, $-\gamma < b < \gamma$ and that the dependence
of $R(\alpha;A,B)$, $y(A,B)$, $h(A,B)$, $F(A,B)$ is
analytical on $A$ and $B$ in this neighborhood. This
assumption is partly warranted by the fact that, if a
solution exists for (\ref{p8}), it develops a pole at $\alpha=
\pm i\gamma$, where $\xi(\alpha)$ has poles.
In the following calculation, the endpoints are kept far away from
the singularities of the inhomogeneous term.
With this assuption, variations can be computed by
taking derivatives in the relevant integral equations.
One starts with (\ref{p8})
\barr
\hspace{-1 cm} \partial_{A}R(\alpha;A,B)
+\frac{1}{2\pi}\int_{A}^{B}d\beta K(\alpha-\beta)
\partial_{A}R(\beta;A,B) &=&
\frac{1}{2\pi}K(\alpha-{A})R(A;A,B) \label{p22} \\
\hspace{-1 cm} \partial_{B}R(\alpha;A,B) 
+\frac{1}{2\pi}\int_{A}^{B}d\beta K(\alpha-\beta) 
\partial_{B}R(\beta;A,B) &=& 
\frac{1}{2\pi}K(\alpha-{B})R(B;A,B) \label{p23} 
\earr
and so forth by taking further derivatives. When $A=A_{0}=
-\pi+ib$ and $B=B_{0}=\pi+ib$, $-\gamma < b < \gamma$ all
these equations can be solved by Fourier transform. For instance
\barrr
\partial_{A}R(\alpha;A_{0},B_{0})&=&\frac{c_{1}}{2\pi}
\sum_{n=-\infty}^{+\infty} e^{-in\alpha}(-)^n
\frac{e^{-nb-\gamma |n|}}{2\cosh n\gamma} =
-\partial_{B}R_{0}(\alpha;A_{0},B_{0}) \\
c_{1} &=& R(A_{0};A_{0},B_{0})=R(B_{0};A_{0},B_{0})
\earrr
Notice that $c_{1}=0$ if $b=\pm \gamma$. This case was dealt
with in \cite{ADW2}. The solution is replaced into the analogous 
expansion for $y(A,B)$, $h(A,B)$ and $F_{R}(A,B)$
(we will consider only the domain where $F_{R}>F_{L}$,
that is $d<\gamma -2u$). As an example, at the first order
\barrr
\frac{\partial y}{\partial A}&=&-\frac{2}{2\pi}\int_{A}^{B}
\partial_{A}R(\alpha;A,B)d\alpha + \frac{2}{2\pi}R(A;A,B) \\
\frac{\partial y}{\partial B}&=&-\frac{2}{2\pi}
\int_{A}^{B}\partial_{B}R(\alpha;A,B)d\alpha - \frac{2}
{2\pi}R(B;A,B)
\earrr
For the field, after using (\ref{p8}) and (\ref{p11})
\barrr
-4i\frac{\partial h}{\partial A}&=& -\frac{1}{2\pi}
\int_{A}^{B}d\beta \partial_{A}R(\beta;A,B)
[\Theta(A-\beta)+\Theta(B-\beta)]+R(A;A,B)[1+\frac{1}{2\pi}
\Theta(B-A)] \\
-4i\frac{\partial h}{\partial B}&=& -\frac{1}{2\pi}
\int_{A}^{B}d\beta \partial_{B}R(\beta;A,B)[\Theta(A-\beta)
+\Theta(B-\beta)]+R(B;A,B)[1+\frac{1}{2\pi}\Theta(B-A)]
\earrr
and finally 
\barrr
\frac{\partial F_{R}}{\partial A} &=& \frac{1}{2\pi}
\int_{A}^{B}d\alpha \partial_{A}R(\alpha;A,B)f_{R}(\alpha;u)
-\frac{1}{2\pi}R(A;A,B)f_{R}(A;u)+\frac{\partial h}{\partial A} \\
\frac{\partial F_{R}}{\partial B} &=& \frac{1}{2\pi}
\int_{A}^{B}d\alpha \partial_{B}R(\alpha;A,B)f_{R}(\alpha;u)
+\frac{1}{2\pi}R(B;A,B)f_{R}(B;u)+\frac{\partial h}{\partial B}
\earrr
To evaluate the last two equations it is useful to know the branch cut
structure of $f_{R}(\alpha;u)$. The cuts run from
$i(\gamma -2u)$ to $+i\infty$ and from $-i(\gamma +2u)$
to $-i\infty$. The transition from the ``$\Lambda_{R}$
regime'' to the ``$\Lambda_{L}$ regime'' occurs when the
integration path crosses the branch cut at $i(\gamma -2u)$,
since $f_{L}(\alpha;u)$ has a branch cut also starting
from $i(\gamma -2u)$ but running all the way down to
$-i\infty$. Furthermore $\mbox{Im}f_{R}(A_{0};u)=\pi$ and
$\mbox{Im}f_{R}(B_{0};u)=-\pi$.

It is now obvious how to go on by taking derivatives. The
expansion has been carried out up to the third order. The
variation of each quantity is a third order polynomial in
$\delta a$, $\delta b$. It is convenient to introduce a more
compact notation that brings out the geometrical meaning
of the coefficients involved. Define
\bd
h_{t}(b)=\frac{d}{db}\overline{h}(b)= -\frac{1}{2}-
\sum_{n=1}^{+\infty}(-)^{n}\frac{\cosh nb}{\cosh n\gamma}=
-\frac{I(k)}{\pi}\mbox{dn}(I+\frac{iIb}{\pi};k)
\ed
where we have used (\ref{p16}) to express the series at hand as
elliptic functions. For $\overline{v}(b)$ one has to 
distinguish between the two cases $b > \gamma -2u$ or $ b < \gamma -2u$, but
the elliptic function expression is the same for both cases
\bd
v_{t}(b)=\frac{d}{db}\overline{v}(b)=\frac{I(k)}{\pi}
\mbox{dn}(I+\frac{iI}{\pi}(2u+b);k)
\ed
We will also need
\barrr
h_{1}(b) &=& \frac{dh_{t}}{db}= ik^2(\frac{I(k)}{\pi})^2
\mbox{sn}(I+\frac{iIb}{\pi};k)\mbox{cn}(I+\frac{iIb}{\pi};k)\\
v_{1}(b) &=& \frac{dv_{t}}{db}=-ik^2(\frac{I(k)}{\pi})^2
\mbox{sn}(I+\frac{iI}{\pi}(b+2u);k)\mbox{cn}
(I+\frac{iI}{\pi}(b+2u);k)
\earrr
$h_{t}$ is negative in the interval $-\gamma < b < \gamma$,
and vanishes when $b=\pm \gamma$. At these two points, the
tangent to $\Gamma$ is parallel to the $v$-axis. Instead, 
$v_{t}=0$ when $b=\gamma-2u$. At this point the tangent to
$\Gamma$ is parallel to the $h$-axis. There is of course 
another point of this kind on the other half of $\Gamma$,
obtained by the symmetry $(h,v) \rightarrow (-h,-v)$. Finally
it is proven in appendix A that the combination $v_{t}h_{1}
-v_{1}h_{t}$, that will appear later, is definite negative.
For polarization and horizontal field one has
\be\label{p24}
\delta y=\frac{h_{t}}{\pi}\delta a -\frac{h_{t}x_{0}}
{2\pi}\delta a^2 + \frac{h_{1}}{\pi}\delta a \delta b +
\frac{1}{2\pi}(\frac{h_{t}x_{0}^2}{2}-\frac{x_{3}}{3})
\delta a^3 - \frac{h_{1}x_{0}}{2\pi}\delta a^2\delta b+
\frac{x_{3}}{2\pi}\delta a\delta b^2
\ee
\be\label{p25}
\delta h=h_{t}\delta b -\frac{h_{1}}{2}\delta a^2 -
\frac{h_{t}x_{0}}{2}\delta a\delta b -\frac{h_{1}x_{0}}{6}
\delta a^3 - \frac{x_{3}}{2}\delta a^2\delta b +\frac{x_{3}}
{6}\delta b^3
\ee
whereas, if we set $Z_0=Z(\gamma -|\gamma -2u-b|)/2\pi$
\barr\label{p26}
\delta F &=& -2h_{t}Z_{0}\delta a +h_{t}x_{0}Z_{0}\delta a^2
-(2h_{1}Z_{0}+\frac{h_{t}v_{t}}{\pi})\delta a\delta b \nonumber \\
&+& [\frac{2h_{1}v_{t}}{\pi}+\frac{h_{t}v_{1}}{\pi}+Z_{0}
(-3h_{t}x_{0}^{2}+2x_{3})]\frac{\delta a^3}{6}+
(\frac{h_{t}v_{t}x_{0}}{2\pi}+h_{1}x_{0}Z_{0})\delta a^2
\delta b \nonumber \\
&-& (\frac{h_{1}v_{t}}{\pi}+\frac{h_{t}v_{1}}{2\pi}
+Z_{0}x_{3})\delta a\delta b^2
\earr
Here $x_{0}$ and $x_{3}$ are non-zero terms that play little role
in what follows
\barrr
x_{0} &=& \frac{1}{\pi}(1+2\sum_{n=1}^{+\infty}\frac{e^{-n\gamma}}
{\cosh n\gamma}) \\
x_{3} &=& -\sum_{n=1}^{+\infty}(-)^{n}n^2\frac{\cosh nb}{\cosh 
n\gamma} = (\frac{I(k)}{\pi})\frac{d^2}{d\alpha^{2}}
\mbox{dn}(\frac{I(k)\alpha}{\pi};k)|_{\alpha=\pm \pi +ib}
\earrr
What actually has to be minimized though is ${\cal F}=-F-vy$.
Using (\ref{p21}) and (\ref{p26})
\barr\label{p27}
\delta {\cal F} &=& -\delta F-\overline{v}(b)\delta y -\delta v
\delta y =-\delta v\delta y +\frac{h_{t}v_{t}}{\pi}\delta a
\delta b \nonumber \\
&-& (\frac{h_{t}v_{1}}{\pi}+\frac{2h_{1}v_{t}}{\pi})\frac{\delta a^3}
{6}+(\frac{h_{1}v_{t}}{\pi}+\frac{h_{t}v_{1}}{2\pi})\delta a
\delta b^2 + \frac{x_{0}v_{t}h_{t}}{2\pi}\delta a^2 \delta b
\earr
These expansions reduce to those of \cite{ADW2} when $b \rightarrow \pm
\gamma$ (the limit should be taken in the elliptic functions because
several series are not convergent when $|b|=\gamma$). No term
$\delta b^{n}$ appears in (\ref{p24}), and that had to be expected since the
line $a=\pi$ (i.e. $\delta a =0$), $-\gamma \leq b \leq \gamma$
corresponds to $y=0$. Furthermore, while (\ref{p8}) and following
equations make perfectly
sense for any values of $A$ and $B$, 
one might object that solutions of (\ref{p6}) can always
be taken to have $-\pi \leq \mbox{Re}\alpha \leq \pi$. Hence $a$ should
vary in $[0,\pi]$. The question at issue is what solutions (\ref{p6})
admit when $n > N/2$, knowing that, when $n=N/2$, they already
fill a line stretching from $-\pi$ to $\pi$. Fortunately, the
question can be bypassed. Only variations $\delta v >0$ will be
considered and it is physically clear that $\delta v >0$ tends to
align arrows ``up'', therefore brings $\delta y >0$, that is
$n < N/2$ and, from (\ref{p24}), $\delta a <0$. This is sufficient because
only the upper half of $\Gamma$, given by (\ref{p17}), (\ref{p21})
is being considered.
The other half, where to drive the system into the incommensurate
phase one needs a variation $\delta v <0$, can, as usual, be 
recovered from $f(h,v)=f(-h,-v)$.

\section{Minimization of ${\cal F}$ and free energy
singularity}
The minimum of ${\cal F}$ should now be taken with respect
to $y$, when $\gamma$, $u$, $h$, $v$ are kept fixed. Keeping
$\delta h$ fixed at a given value means that $\delta a$ and
$\delta b$ are not independent. Two different ways will be followed
to deal with (\ref{p24}),(\ref{p25})and (\ref{p27}) and they give the same 
results.

Neglecting terms $\da^3$, $\db^3$ in (\ref{p25}) and solving for $\db$
one finds
\be\label{p28}
\db = \frac{\dh}{h_{t}}-\frac{h_{1}}{2h_{t}}\da^2 - \frac{h_{1}}
{2h_{t}^3}\dh^2 + \frac{x_{0}}{2h_{t}}\da \dh + O(\da^3,\da^2\dh,
\da \dh^2,dh^3)
\ee
when inserted into $\delta {\cal F}$, one gets
\barr
\delta {\cal F} &=& \da (-\frac{h_{t}}{\pi}\dv -\frac{h_{1}}{h_{t}\pi}\dv \dh
+ \frac{v_{t}}{\pi}\dh +\frac{v_{1}h_{t}+h_{1}v_{t}}{2\pi h_{t}^{2}}\dh^2
+ \ldots) \nonumber \\
&+& \da^2 (\frac{h_{t}x_{0}}{2\pi}\dv + c_{2} \dh^2 +\ldots) + \frac{\da^3}
{6\pi}\Delta +\ldots \label{p29} \\
\Delta &=& v_{t}h_{1}-v_{1}h_{t} \label{p30}
\earr
All terms neglected are higer order, i.e. $\da^4,\da^3\dh,\da^2\dv\dh^2,
\da \dv \dh^2$, etc. and it will soon become clear that dropping
them is justified. The coefficient $c_{2}$ is $b$-dependent and to know
its specific value will not be necessary in the following.

Let's consider the coefficient of $\da$ in (\ref{p29}). It is easy to see, 
from (\ref{p17}) and (\ref{p21}) that it vanishes when the variation 
$(\dh,dv)$ is taken
along $\Gamma$. This had to be expected. The curve $\Gamma$ is described
by minima of ${\cal F}$ falling on the line $a=\pi$, $-\gamma \leq b \leq
\gamma$ or, stated otherwise, with $\da=0$. In fact, the vanishing of the 
first order
term in (\ref{p29}) implies that $\delta {\cal F}$ has a stationary point at
$\da=0$. Suppose next that we approach $\Gamma$ along any direction
other than the tangential one, that is
\be\label{p31}
\dv = k\dh \ \ \ \ k \neq \frac{v_{t}}{h_{t}}
\ee
To move into the incommensurate phase one has
to take $\dh > 0$ for $k>v_{t}/h_{t}$ and $\dh <0$ for $k<v_{t}/h_{t}$.
Clearly the linear terms in the coefficient of $\da$ in (\ref{p29})
do not cancel and
therefore the terms $\dv\dh, \dh^2$ (and higher) in the same coefficient
can be dropped. Same thing can be said about the $\da^2$ term since,
obviously, $\da^2<<\da$. The term $\da^3$ must be retained in all cases,
because its coefficient is finite. Instead terms like $\da^3\dh$, $\da^4$
etc. are clearly negligible compared with the $\da^3$ term. It can be
seen by inspection that the terms thrown away in solving (\ref{p25}) are also
negligible. The whole procedure assumes though that a small
variation $(\dh,\dv)$ brings about a small displacement $(\da,\db)$
in the minimum of ${\cal F}$. The upshot is that it is legitimate to
keep
\be\label{p32}
\delta {\cal F}= \frac{\da^3}{6\pi}\Delta + \da(-\frac{h_{t}}{\pi}\dv
+\frac{v_{t}}{\pi}\dh)
\ee
Once (\ref{p31}) is taken into account, the solution of
\bd
\frac{\partial \delta{\cal F}}{\partial \da}=0
\ed
occurs at
\be\label{p33}
\da_{0}=-\left[\frac{2(kh_{t}-v_{t})\dh}{\Delta}\right]^{1/2}
\ee
Notice that the previous limits on $k$ guarantee that the solution
is real. The sign has been chosen to make sure that $\da<0$. An
elementary check shows that $\partial^{2}(\delta {\cal F})/\partial
(\da)^2|_{\da_{0}}>0$ and it confirms that $\da_{0}$ is a minimum.
When $\da_{0}$ is plugged into (\ref{p29}) one finds that, 
when $\dv=k\dh$, $k \neq \frac{v_{t}}{h_{t}}$
\be\label{p34}
\delta f=f(\gamma,u,h+\dh,v+\dv)-f_{0}(\gamma,u)=
(\frac{h_{t}}{\Delta})^{1/2}\frac{h_{t}}{3\pi}[2(k-\frac{v_{t}}{h_{t}})
\dh]^{3/2}
\ee
So the exponent is $3/2$ for all these directions. We pass next to
the case where the transition line is approached tangentially, i.e.
\bd
\dv=\frac{v_{t}}{h_{t}}\dh
\ed
We have then
\be\label{p35}
\delta {\cal F}=\frac{\da^3}{6\pi}\Delta + \frac{v_{t}x_{0}}{2\pi}
\dh\da^{2}-\da\dh^{2}\frac{\Delta}{2\pi h_{t}^{2}}
\ee
The minimum lies at
\be\label{p36}
\da_{0}=-\left[v_{t}x_{0}\dh +\frac{|\dh|}{h_t}\sqrt{\Delta^2+(v_{t}h_{t}
x_{0})^2}\right]/\Delta
\ee
Since $\partial^{2}(\delta {\cal F})/\partial (\da)^{2}|_{\da_{0}}>0$,
$\da_{0}$ is indeed a minimum. Inserting it into $\delta {\cal F}$
yields a jump in the third derivative of the free energy. Namely, at
$h_{t}\dv=v_{t}\dh$
\be\label{p37}
\delta f=f(\gamma,u,h+\dh,v+\dv)-f_{0}(\gamma,u)=
\left\{ \begin{array}{l} c_{+}\dh^3 \ \ \mbox{if}\ \ \dh>0 \\
c_{-}\dh^3 \ \ \mbox{if}\ \ \dh<0 \end{array} \right.
\ee
where
\be\label{p38}
c_{\pm}=\frac{1}{6\pi h_{t}^{3}\Delta^{2}}(g \pm \sqrt{g^2+\Delta^2})
(g^2+2\Delta^2 \pm g\sqrt{g^2+\Delta^2}) \ \ \ g=v_{t}x_{t}x_{0}
\ee
When $2u+b=\gamma$, $v_{t}=0$. Then (\ref{p38}) somewhat simplifies, since
$g=0$ and the variation tangential to $\Gamma$ is just $\dh$, so
\bd
\delta f=-\frac{|v_{1}|}{3\pi h_{t}^2}|\dh|^3
\ed
An analogous result, with $\dv \leftrightarrow \dh$ was reached in 
\cite{ADW2}, but only for the two points $b=\pm \gamma$, where the tangent 
to $\Gamma$ is parallel to the $v$-axis

A second method of calculation will now be presented. From (\ref{p15}),
the equation determining the equilibrium value of $y$ is
\be\label{p39}
\frac{\partial F}{\partial y}|_{h}+v=0
\ee
Eq. (\ref{p20}) is just a particular case of this condition. It identifies
the value of $v$ for which the minimum occurs at $y=0$. If one thinks
to invert (\ref{p15}), (\ref{p24}), $F$ becomes a function of $a$, $b$, 
so \cite{No}
\be\label{p40}
-v=\frac{\partial F}{\partial a}|_{b}\frac{\partial a}{\partial y}|_{h}+
\frac{\partial F}{\partial b}|_{a}\frac{\partial b}{\partial y}|_{h}=
\frac{\frac{\partial F}{\partial b}\frac{\partial h}{\partial a}-
\frac{\partial F}{\partial a}\frac{\partial h}{\partial b}}
{\frac{\partial h}{\partial a}\frac{\partial y}{\partial b}-
\frac{\partial h}{\partial b}\frac{\partial y}{\partial a}}
\ee
which makes sense if (\ref{p24}) and (\ref{p25}) are indeed invertible, 
that is if the
denominator in the RHS does not vanish. 
It is immediate to see from (\ref{p24}), (\ref{p25}) that this
is true for the points we are considering. Instead, the points $b=\pm \gamma
(h_{t}=0)$ are saddle points for $h(a,b)$ \cite{ADW2}. 
Expanding the denominator
in (\ref{p40}) and retaining terms up to the second order ( to have the third
order in (\ref{p40}) one should expand $y$, $h$ and $F$ to the fourth order)
\bd
-v=-2\pi Z_{0}-v_{t}\db+\frac{v_{1}}{2}\da^2-\frac{v_{1}}{2}\db^2
-\frac{x_{0}v_{t}}{2}\da \db
\ed
Writing $v=\overline{v}(b)+\dv$, owing to (\ref{p21}) one arrives at
\barr
\dv &=& v_{t}\db -\frac{v_{1}}{2}(\da^2 - \db^2) + \frac{v_{t}x_{0}}{2}
\da \db + \ldots \label{p41} \\
\dh &=& h_{t}\db -\frac{h_{1}}{2}(\da^2 - \db^2) - \frac{h_{t}x_{0}}{2}
\da \db +\ldots \label{p42}
\earr
whose solution yields $\da(\dh,\dv)$ and $\db(\dh,\dv)$. Their result
has to be used in (\ref{p29}) to produce the leading singular part of 
$\delta f$.
Again, two cases have to be distinguished. If $\dv=k\dh$, $k \neq 
v_{t}/h_{t}$, consistency of (\ref{p40}), (\ref{p41}) requires 
that $\db \sim c\da^2 +
o(\da^2)$, so at the leading order
\barrr
\dv &=& (cv_{t}-\frac{v_{1}}{2})\da^2 \\
\dh &=& (ch_{t}-\frac{h_{1}}{2})\da^2
\earrr
Hence
\bd
c=\frac{1}{2}\frac{v_{1}-kh_{1}}{v_{t}-kh_{t}} \ \ \
\da=-\left[\frac{2(kh_{t}-v_{t})\dh}{\Delta}\right]^{1/2}
\ed
which coincides with (\ref{p33}). $\delta {\cal F}$ is obviously the same
previously used, so the free energy coincides with (\ref{p34}). Suppose
instead $\dv=k\dh$ but $k=v_{t}/h_{t}$. Clearly (\ref{p41}), (\ref{p42}) 
give $\db=\dh/h_{t}$ at the leading order and
\bd
\Delta \da^2 +2v_{t}x_{0}\da\dh - \Delta \frac{\dh^2}{h_{t}^2}=0
\ed
whose solution is, again (\ref{p36}). Yet, replacing $\db=\dh/h_{t}$ and
(\ref{p36}) into (\ref{p27}) is not correct. 
The reason is that a term $\da\db$
appears in (\ref{p27}) that would require solving (\ref{p41}) and (\ref{p42})
for $\db$ up to the next
order in $\dh$. The problem can be bypassed by replacing e.g. (\ref{p42}) 
into the term $\da\db$ of (\ref{p27}). So one gets
\bd
\delta {\cal F}=\frac{\da^3}{6\pi}\Delta +\frac{x_{0}v_{t}}{2\pi}
\dh\da^2-\frac{v_{t}h_{1}}{\pi h_{t}}\dh\da\db+\frac{\da\db^2}{2\pi}
(h_{1}v_{t}+h_{t}v_{1})
\ed
and after setting $\db=\dh/h_{t}$ one is back to (\ref{p35}) and, because of
(\ref{p36}), to the $\delta f$ of (\ref{p37}).

\section*{Acknowledgements}
The author is particularly grateful to the organizers of the
``Integrable semester'' at the Institut Henri Poincare' in Paris where part
of this work was carried out. Useful discussions with dr.S.Dahmen,
prof.B.M.McCoy, prof.V.Rittenberg, prof.C.Viallet and B.Wehefritz
are also acknowledged.

\appendix
\section{Appendix}
One has to prove that $\Delta<0$ for all $b$ in $(-\gamma,\gamma)$.
Setting
\bd
x=I+\frac{iIb}{\pi} \hspace{3cm} y=I+\frac{iI}{\pi}(b+2u)
\ed
from (\ref{p30})
\bd
\Delta=(\frac{I}{\pi})^{3}ik^{2}\mbox{dn}(x;k)\mbox{dn}(y;k)\,
\left[\frac{\mbox{sn}(x;k)\mbox{cn}(x;k)}{\mbox{dn}(x;k)}-
\frac{\mbox{sn}(y;k)\mbox{cn}(y;k)}{\mbox{dn}(y;k)}\right]
\ed
Perform Landen's trasform \cite{Er}
\barrr
k \rightarrow k_{1} &=& \frac{1-k'}{1+k'} \hspace{3cm} u \rightarrow 
u_{1}=(1+k')u \\
\mbox{sn}(u_{1};k_{1}) &=& \frac{k}{k_{1}^{1/2}}\, 
\frac{\mbox{sn}(u;k)\mbox{cn}(u;k)}{\mbox{dn}(u;k)}
\earrr
The new half-periods are
\bd
I_{1}=I(k_{1})=\frac{1+k'}{2}I(k) \hspace{3cm}
I_{1}'=I'(k_{1})=(1+k')I'(k)
\ed
and so
\bd
\Delta=k_{1}^{1/2}ik(\frac{I}{\pi})^3 \mbox{dn}(x;k)\mbox{dn}(y;k)\,
[\mbox{sn}(\frac{iI_{1}'}{\gamma}(b+2u);k_{1})-\mbox{sn}(\frac{iI_{1}'b}
{\gamma};k_{1})]
\ed
A table of signs of $\mbox{dn}$, $\mbox{cn}$ and $\mbox{sn}$ is given
in \cite{Er}. By inspecting the possible cases one sees that $\Delta<0$ 
always. A possible exception comes from $\mbox{dn}(y;k)=0$ that is when
$b+2u=\gamma$. In this case $v_{t}=0$ and 
\bd
\Delta=-v_{1}h_{t}(b=\gamma -2u)=h_{t}k'(\frac{I}{\pi})^2<0
\ed
There might be a simpler proof.

\newpage
%
%
%
%
\begin{figure}[h]
\setlength{\unitlength}{5mm}
\begin{picture}(54,24)
\put(0,0){\framebox(32,24)}
\put(3,1){\vector(0,1){1}}
\put(3,3){\vector(0,-1){1}}
\put(3,2){\vector(1,0){1}}
\put(3,2){\vector(-1,0){1}}
\put(3,6){\vector(0,1){1}}
\put(3,6){\vector(0,-1){1}}
\put(4,6){\vector(-1,0){1}}
\put(2,6){\vector(1,0){1}}
\put(3,10){\vector(0,-1){1}}
\put(3,11){\vector(0,-1){1}}
\put(3,10){\vector(1,0){1}}
\put(2,10){\vector(1,0){1}}
\put(3,13){\vector(0,1){1}}
\put(3,14){\vector(0,1){1}}
\put(3,14){\vector(-1,0){1}}
\put(4,14){\vector(-1,0){1}}
\put(3,18){\vector(0,-1){1}}
\put(3,19){\vector(0,-1){1}}
\put(3,18){\vector(-1,0){1}}
\put(4,18){\vector(-1,0){1}}
\put(3,21){\vector(0,1){1}}
\put(3,22){\vector(0,1){1}}
\put(2,22){\vector(1,0){1}}
\put(3,22){\vector(1,0){1}}
\put (6,1.8) {=}
\put (6,5.8) {=}
\put (6,9.8) {=}
\put (6,13.8) {=}
\put (6,17.8) {=}
\put (6,21.8) {=}
\put (8,1.8) {1}
\put (8,5.8) {1}
\put (8,9.8) {$e^{h-v}\;\frac{\sinh u}{\sinh \gamma} $}
\put (8,13.8) {$e^{-h+v}\;\frac{\sinh u}{\sinh \gamma}$}
\put (8,17.8) {$e^{-h-v}\;\frac{\sinh(\gamma-u)}{\sinh \gamma}$}
\put (8,21.8) {$e^{h+v}\;\frac{\sinh(\gamma-u)}{\sinh \gamma}$ }
\put (13,1.8) {=}
\put (13,5.8) {=}
\put (13,9.8) {=}
\put (13,13.8) {=}
\put (13,17.8) {=}
\put (13,21.8) {=}
\put (15,1.8) {1}
\put (15,5.8) {1}
\put (15,9.8)  {$ e^{\beta (-\delta/2 - \epsilon) + h - v} $}
\put (15,13.8) {$ e^{\beta (-\delta/2 - \epsilon) - h + v} $}
\put (15,17.8) {$ e^{\beta (\delta/2 - \epsilon) - h - v}$}
\put (15,21.8) {$ e^{\beta (\delta/2 - \epsilon) + h + v}$ }
\put (21,1.8) {=}
\put (21,5.8) {=}
\put (21,9.8) {=}
\put (21,13.8) {=}
\put (21,17.8) {=}
\put (21,21.8) {=}
\put (23,1.8) {$R^{21}_{12} (u)$}
\put (23,5.8) {$R^{12}_{21} (u)$}
\put (23,9.8) {$R^{11}_{22} (u)$}
\put (23,13.8) {$R^{22}_{11} (u)$}
\put (23,17.8) {$R^{22}_{22} (u)$}
\put (23,21.8) {$R^{11}_{11} (u)$}
\put (1.5,1.8) {{\scriptsize 2}}
\put (2.9,0.5) {{\scriptsize 1}}
\put (2.9,3.3) {{\scriptsize 2}}
\put (4.3,1.8) {{\scriptsize 1}}
\put (1.5,5.8) {{\scriptsize 1}}
\put (2.9,4.5) {{\scriptsize 2}}
\put (2.9,7.3) {{\scriptsize 1}}
\put (4.3,5.8) {{\scriptsize 2}}
\put (1.5,9.8) {{\scriptsize 1}}
\put (2.9,8.5) {{\scriptsize 2}}
\put (2.9,11.3) {{\scriptsize 2}}
\put (4.3,9.8) {{\scriptsize 1}}
\put (1.5,13.8) {{\scriptsize 2}}
\put (2.9,12.5) {{\scriptsize 1}}
\put (2.9,15.3) {{\scriptsize 1}}
\put (4.3,13.8) {{\scriptsize 2}}
\put (1.5,17.8) {{\scriptsize 2}}
\put (2.9,16.5) {{\scriptsize 2}}
\put (2.9,19.3) {{\scriptsize 2}}
\put (4.3,17.8) {{\scriptsize 2}}
\put (1.5,21.8) {{\scriptsize 1}}
\put (2.9,20.5) {{\scriptsize 1}}
\put (2.9,23.3) {{\scriptsize 1}}
\put (4.3,21.8) {{\scriptsize 1}}
\end{picture}
\caption{Boltzmann weights in the notation with spectral parameter $u$ 
compared to that of ref.[2]. The physical region is $0 < u < \gamma$} 
\label{picture}
\end{figure}
\end{document}